\documentclass[12pt]{article}

\usepackage{amsmath,amssymb,epsfig}
\usepackage{graphicx}

\begin{document}
\def\refname{\Large~~~~~~~{\bf References}}
\newcommand{\el}{\left}
\newcommand{\er}{\right}
\newcommand{\p}{\prime}
\newcommand{\ro}{\rho^\circ}
\newcommand{\ti}{\tilde}
\newcommand{\veps}{\varepsilon}
\newcommand{\dis}{\displaystyle}
\newcommand{\scr}{\scriptsize}
\begin{center}
{\Large \bf HIGH-ENERGY APPROACH FOR HEAVY-ION \\
%
SCATTERING WITH EXCITATIONS OF\\

\vspace*{2mm} NUCLEAR COLLECTIVE STATES
}\\[5mm]

{\large\bf  V.K.~Lukyanov$^1$, Z.~Metawei$^2$,
E.V.~Zemlyanaya$^1$}\\[3mm]
{\large\it $^1$Joint Institute for Nuclear Research, Dubna, Russia~~~~~~~~

$^2$Physics Department, Faculty of Science, Cairo University,\\ Giza, Egypt}\\
\vspace{0.5cm}
\end{center}

{\small A phenomenological optical potential is generalized to
include the Coulomb and nuclear interactions caused by the
dynamical deformation of its surface. In the high-energy approach
analytical expressions for elastic and inelastic scattering
amplitudes are obtained where all the orders in the deformation
parameters are included. The multistep effect of the 2$^+$
rotational state excitation on elastic scattering is analyzed.
Calculations of inelastic cross sections for the $^{17}$O ions
scattered on different nuclei at about hundred Mev/nucleon are
compared with experimental data, and important role of the Coulomb
excitation is established. }
\vspace{0.5cm}

{\normalsize
\section {Introduction}

Diffraction theory of excitations of collective states of nuclei by
scattered nucleons was firstly considered in \cite{D1},\cite{I1},\cite{B1},
where the adiabatic approximation for the amplitude of a process was used.
In the case of excitation of the low lying rotational or vibrational
state $|IM>$ of the even-even nuclei, having the ground state spin and its
projection $|00>$, this amplitude is
\begin{equation}\label{c1}
f_{IM}(q)\, = \,<IM|\,f(q,\{\alpha_{\lambda\mu}\})|00>,
\end{equation}
where $q=2k\sin(\vartheta/2)$ is the transfer momentum, $k$ is the
relative momentum, and $\vartheta$, the angle of scattering. Here
$f(q,\{\alpha_{\lambda\mu}\})$ is the amplitude of elastic
scattering on a system with the "frozen" coordinates of collective
motion $\{\alpha_{\lambda\mu}\}$. These latter are introduced with
a help of the radius parameter
\begin{equation}\label{c2}
\Re\, =\,R\,+\,\delta R, \qquad
\delta R\,=\,R\sum\limits_{\lambda\mu}\,\alpha_{\lambda\mu}\,
Y_{\lambda\mu}(\theta,\phi),
\end{equation}
 where $\theta,\phi$ are spherical coordinates of the space
vector ${\bf r}$ in the laboratory system. In
\cite{D1},\cite{I1},\cite{B1}, the diffraction theory was applied
to obtain (\ref{c1}) in the certain form using the first order
expansion in the value of $\delta R$.

Later on, e.g., in \cite{Ber1},\cite{Star1}, in the framework of the
Glauber-Sitenko microscopic diffraction theory \cite{Gla},\cite{Sit}, the
amplitude $f(q,\{\alpha_{\lambda\mu}\})$ was derived including
the second order terms in the value of $\delta R$,
and comparisons with experimental data were made for the proton scattering
from different nuclei with excitations of the 2$^+$, 3$^-$ collective states.
Besides, general methods of accounting for higher approximations with respect to
the deformations $\{\alpha_{\lambda\mu}\}$ were presented in \cite{LP},
\cite{ISh},\cite{FG}, but no applications to analysis of experimental data
were made. In all the above mentioned works the Coulomb forces in collisions
were not included into considerations.

In this paper, we consider effects of the virtual excitation of the
target-nucleus rotational 2$^+$ state on the heavy-ion elastic scattering,
and also we calculate the respective inelastic differential cross-sections
with excitation of this state. In the case of nucleus-nucleus scattering,
there are specific features, namely, strong absorption in the inner region
of an interaction, the high sensitivity of scattering to the shape of
potentials in peripheral region sending us to search the realistic, with the
exponential slope, Fermi-like form of a potential, and also the strong Coulomb
potential may not be excluded from considerations.

We start with the expression for the elastic scattering amplitude in the
high-energy approximation (HEA) using a phenomenological nucleus-nucleus
optical potential:
\begin{equation}\label{c3}
f(q)\, =\, i{k\over 2\pi} \int \, bdbd\phi \, {\dis e}^{\dis iqb\cos\phi}
\, \Bigl [1-{\dis e}^{\dis i\Phi}\Bigr ].
\end{equation}
Here integration is performed over impact parameters $b$ and on its
azimuthal angle $\phi$.
The eikonal phase is determined by a potential of scattering as follows:
\begin{equation}\label{c4}
\Phi(b)\,=\,-{1\over \hbar v}\,\int_{-\infty}^\infty U({\bf r})\,dz,
\qquad r=\sqrt{b^2+z^2},
\end{equation}
 where $v$ is the relative velocity of colliding nuclei. In
general, the potential includes dependence on angles $\theta,\phi$
of vector ${\bf r}$ in the laboratory system and on coordinates
$\alpha_{\lambda\mu}$ responsible for intrinsic collective motion.
Note, that in (\ref{c4}) the polar angle $\phi$ of vector ${\bf
r}$ coincides with that in eq.(\ref{c3}) where the cylindrical
frame is used. Also, in this frame the angle $\theta$ is defined
by $\cos\theta=z/\sqrt{b^2+z^2}$.

The expression (\ref{c3}) is valid for $E\gg |U|$ and at small scattering
angles $\vartheta < \sqrt{2/kR}$ with $R$, the radius of a potential
(see e.g. \cite{LZ2}).

Usually, to get the deformed part of the potential, one
substitutes the deformed radius (\ref{c2}) instead of $R$ and then
expands the spherically symmetrical optical potential in the value
of $\delta R/R$. As the result, the phase consists of the central
and deformed parts:
\begin{equation}\label{c5}
\Phi(b,\Re)=\Phi_0(b)\,+\,\Phi_{int}(b,\, \{\alpha_{\lambda\mu}\},\,\phi).
\end{equation}

Furthermore, for scattering on even-even nuclei, we consider the deformed
axially symmetrical quadrupole ($\lambda$=2) optical potential and excitations
of the $2^+$ rotational state. Then, the rotational wave functions and
collective variables are as follows
\begin{equation}\label{c6}
|I\,M>=\sqrt{2I+1\over 8\pi^2}\,D^{(I)}_{M0}(\Theta_i), \qquad
\alpha_{2\,\mu}\,=\,\beta_2 \,D^{(2)\,*}_{\mu 0}(\Theta_i),
\end{equation}
where $\beta_2$ is the static deformation parameter, and $\Theta_i$
are rotational angles.
In this case, it was shown in \cite{B1} and \cite{ISh} that amplitudes
with only even projections $M=0,\pm 2$ contribute to the cross-sections
and that there exists the equality $f_{2\,-2}(q)=f_{2\,2}(q)$. Therefore,
elastic and inelastic differential cross sections are expressed as follows
\begin{equation}\label{c7}
{d\sigma_{el}\over d\Omega}\,=\,|f_{0\,0}(q)|^2,\qquad
{d\sigma_{in}\over d\Omega}\,=\,|f_{2\,0}|^2\,+\,2\,|f_{2\,2}(q)|^2.
\end{equation}

\section {Transition potentials and phases}

In nuclear physics, the Woods-Saxon shape is usually used as a
form of a phenomenological nuclear potential. In our calculations
we apply its symmetrical form (symmetrical Fermi function) for the
central part of optical potential
\begin{equation}\label{d8}
U_0^{(N)}=(V_0+iW_0)\,u_{SF}(r, R),
\end{equation}
\begin{equation}\label{d9}
u_{SF}(r, R)\,=\,{\sinh(R /a)\over \cosh(R /a)+\cosh(r/a)}\simeq
{1\over{1+\exp{r-R\over a}}},
\end{equation}
where $R=r_0(A^{1/3}_1+A^{1/3}_2)$. Its quadrupole part is
obtained by exchanging $R$ by $\Re$ and by leaving the first order
term of its expansion in the $\delta R$, namely,
\begin{equation}\label{d10}
U^{(N)}_{int}\,=\,(V_0\,+\,iW_0)\,R\,{d\over dR}\,u_{SF}(r,R)
\sum\limits_\mu\,\alpha_{2\mu} \, Y_{2\mu}(\theta,\phi).
\end{equation}
The Coulomb part of the nucleus-nucleus potential is obtained
using the definition
\begin{equation}\label{d11}
U^{(C)}\,=\,Z_1e\,\int {\rho({\bf r}^\prime)\over |{\bf r}\,-
\,{\bf r}^\prime |}\,d^3r^\prime
\end{equation}
As usually, we take here the uniform density distribution depended
of deformation, $\rho({\bf
r},\,\Re)=\rho_0[\Theta(R-r)+\delta(R-r)\delta R(\theta,\phi)]$,
where ~$\rho_0=3Z_2e/4\pi R_C^3$~, ~$R_C=r_c(A_1^{1/3}+
A_2^{1/3})$, and the step function ~$\Theta(x)$~ is equal to 1 for
~$x>0$~ and 0 for ~$x\leq 0$. For the spherically symmetrical part
of the density $\rho_0\Theta(R-r)$ one obtains the known
expression for the central Coulomb potential $U^{(C)}_0(r)$ and
the corresponding phase $\Phi^{(C)}_0(b)$ (see, e.g., \cite{LZ2}).
Furthermore, using the second term $\rho_0\delta(R-r) \delta
R(\theta,\phi)$ the quadrupole part can be derived from
(\ref{d11}) as follows
\begin{equation}\label{d12}
U^{(C)}_{int}\,=\,{3\over 5}\,U_B\,
\Bigl[\Bigl({r\over R_C}\Bigr)^2\Theta(R-r)+\Bigl({R_C\over r}\Bigr)^3
\Theta(r-R)\Bigr]\,\sum\limits_\mu\, \alpha_{2\mu}
\,Y_{2\mu}(\theta,\phi),
\end{equation}
where ~$U_B=Z_1Z_2e^2/R_C$.

It is known that in collisions of two nuclei the outer region $\bar r\simeq
R_1+R_2+a$ plays the determined role in direct reactions (see, e.g.,
\cite{LSZ}). So, estimating strengths of the nuclear and Coulomb interaction
potentials (\ref{d10}) and (\ref{d12}) at $\bar r$ in the case, for example,
of inelastic scattering of ${^{17}}$O+${^{90}}$Zr, one finds them to be
of the same order of the values. Thus, when developing a theory of the
heavy ion inelastic scattering, the Coulomb excitation may not be omitted
as this usually made in the case of the proton and alpha-particle projectiles.

Now, substituting (\ref{d10}) and (\ref{d12}) in (\ref{c4}), and
using $\alpha_{2\,\mu}$ from (\ref{c7}) and equality
$Y_{2,\mu}(\theta,\phi)= Y_{2\mu}(\theta,0){\dis e}^{\dis
i\mu\phi}$ one can get the transition phase in (\ref{c5})
\begin{equation}\label{d13}
\Phi_{int}\,=\,\beta_2\sum\limits_{\mu=0,\pm 2}\, G_{\mu}(b)
\,D^{(2)\,*}_{\mu\,0}(\Theta_i)\,
{\dis e}^{\dis i\mu\phi}, \qquad
G_{\mu}(b)\,=\,G^{(N)}_{\mu}(b)\,+\,G^{(C)}_{\mu}(b).
\end{equation}
 Here
\begin{equation}\label{d14}
G^{(N)}_{\mu}\,=\,-{2\over \hbar v}\,(V_0\,+\,iW_0)\,R\,
\int_0^\infty dz\,{du_{SF}(r,R)\over dR}Y_{2\mu}
(\arccos(z/r),0), \quad  r=\sqrt{b^2+z^2},
\end{equation}
and the Coulomb quadrupole phases have the certain analytical forms
%
\begin{equation}\label{d15}
G^{(C)}_{\mu=0}\,=\,-{1\over \hbar v}
{1\over 2}\,\sqrt{5\over\pi}\,
U_B\,R_C\,\el(1\,-\,{b^2\over R^2_C}\er)^{3/2}\,\Theta(R_c-b)\,
\end{equation}
%
$$
G^{(C)}_{\mu=2}\,=\,-{1\over \hbar v}\,\sqrt{3\over 10\pi}\,U_B\,R_C\,\times
~~~~~~~~~~~~~~~~~~~~~~~~~~~~~~
$$
\begin{equation}\label{d16}
\el\{{R_C^2\over b^2}\,\Theta(b-R_C)\,+\,\el[{R_C^2\over b^2}
\Biggl(1-\sqrt{1-{b^2\over R^2_C}}\,\Biggr)\,-\,\sqrt{1-{b^2\over R^2_C}}\,\er]
\Theta(R_C-b)\er\}
\end{equation}

\section {Numerical calculations and discussion}
\subsection {Effect of deformation on elastic scattering}

The deformation effect on elastic scattering can be explicitly
calculated if one leaves only $\mu$=0 term in the quadrupole phase
(\ref{d13}). Indeed, in this case, the phase $\Phi_{int}$ does not
depend on $\phi$ and also $D^{(2)\,*}_{0\,0}(0,\beta,0)= P_2(x)$,
where $x=\cos\beta$. Then integration over $\phi$ in (\ref{c3}) is
performed resulting to the Bessel function, and integrations in
(\ref{c1}) on rotational angles $\Theta_i=\{\alpha,\beta,\gamma\}$
reduce to the one-dimensional integral
\begin{equation}\label{e17}
f^{(\mu=0)}_{0\,0}(q)\,= \,ik\, \int_0^\infty \, bdb\, J_0(qb)\,
\Bigl [1-{\dis e}^{\dis i\Phi_0}\,E_0(b)\Bigr],
\end{equation}
where
\begin{equation}\label{e18}
E_0\,=\,\int_0^1\,dx\, {\dis e}^{\dis i\beta_2\,G_0(b)\,P_2(x)}
\end{equation}
When expanding the exponential function in (\ref{e18}) we find
that the deformation admixture to elastic scattering amplitude
arrives beginning from the second power term of the deformation
parameter $\beta_2$, because of equalities  $\int_0^1P_2(x)dx$=0
and $\int_0^1[P_2(x)]^2dx$=1/5. However, if we limit ourselves by
only the second power of $\beta_2$, then, in the scattering
amplitude, the $\mu=2$ terms can be also taken into account, which
gives
%
\begin{equation}\label{e19}
f^{(\mu=0,2)}_{0\,0}(q)\,= \,ik\, \int_0^\infty \, bdb\, J_0(qb)\,
\Bigl [1-{\dis e}^{\dis i\Phi_0}\Bigr]\,+ ~~~~~~~~~~~~~~~~~~~~~~~~~~~~~~~~~~~~
\end{equation}
$$
~~~~~~~~~~~~~~~~~~~~~~
+\,ik(\beta_2)^2{1\over 10}\,\int_0^\infty \, bdb\,\el[ J_0(qb)\,
G_0^2(b)\,+\,2\,J_4(qb)\,G_2^2(b)\er]\,{\dis e}^{\dis i\Phi_0}.
$$
Note that in calculations of both elastic and inelastic
differential cross sections one should take into account the
Coulomb trajectory distortion. To this end the prescription is
usually utilizes when in the nuclear part of the phase $\Phi_0$,
the impact parameter $b$ is exchanged by the distance to the
turning point
\begin{equation}\label{e20}
b_c\,=\,\bar a\,+\,\sqrt{b^2+{\bar a}^2},
\end{equation}
where $\bar a=Z_1Z_2e^2/\hbar vk$ is a half-distance of the closest
approach in the Coulomb field at $b$=0.

\begin{figure}[t]
\begin{center}
\includegraphics[ height = 3.in, width =
.95\linewidth]{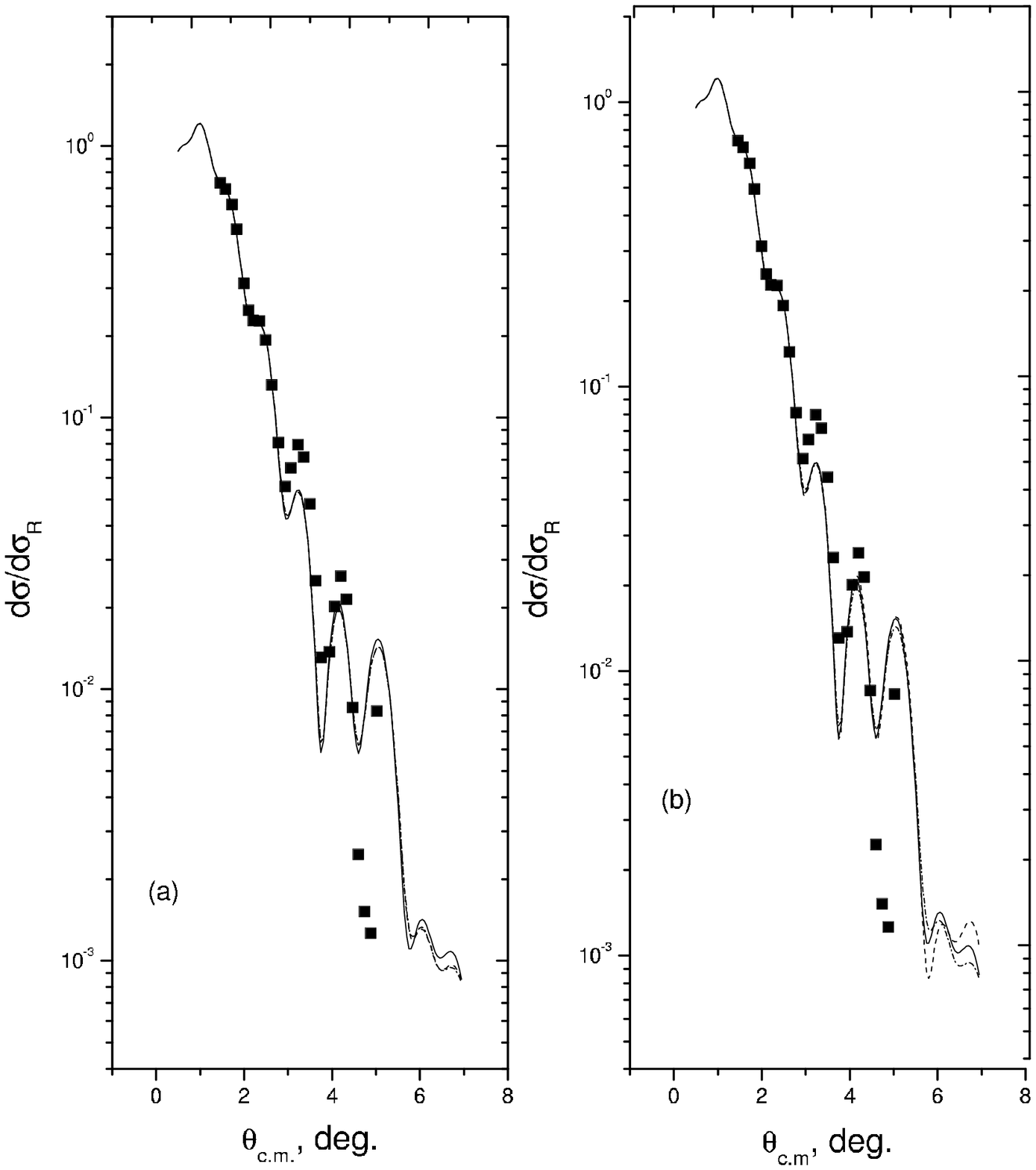}
\end{center}
{Fig. 1. The ratio of the elastic scattering differential cross
sections of ${^{17}}$0+${^{60}}$Ni at E$_{lab}$=1435 MeV to the
Rutherford one. Experimental data are from \cite{Neto}, solid
curves - calculations without deformation effect. (a), case
$\mu=0$: dot-dashed curve - deformation is included rigorously in
(\ref{e17}); dashed curve - only the zero and the second power
terms of $\beta_2$ in (\ref{e18}) are included. (b),  dot-dashed
curve - deformation is included rigorously in (\ref{e17}) when
$\mu=0$, dashed curve - calculations with $\mu=0,\,\pm2$ terms
included in the second power of $\beta_2$ by (\ref{e19}). }
\end{figure}

Figure 1 shows the ratio of differential cross sections of elastic scattering
of ${^{17}}$0+${^{60}}$Ni at E$_{lab}$=1435 MeV, calculated with a help of
(\ref{e17}),(\ref{e18}), to the Rutherford cross section. Here and below we
use the experimental data and parameters of the spherically symmetrical optical
potential from \cite{Neto} and take the deformation parameter $\beta_2=0.236$.
Solid curves correspond to the case when there is no deformation effect
($\beta_2$=0, E$_0$=1). The dot-dashed curve in Fig.1(a) illustrates
calculations when $\mu=0$ and the deformation effect is included rigorously
by computing (\ref{e17}), whereas the dashed curve is the case when one takes
in (\ref{e18}) only terms of the zero and the second power of $\beta_2$.
In Fig.1(b), the dot-dashed curve exhibits effect of both $\mu=0$ and
$\mu=\pm 2$ terms of the second power contributions
in (\ref{e19}). The noticeably deformation effect is seen only at fairly
large angles of scattering. Also, contributions of the higher power terms
of $(\beta_2)^n$ ($n\geq 3$) occur small one, and, in elastic scattering, one
can leave in calculations only terms of the zero and the second order in
$\beta_2$. Actually, these results have the methodical meaning, useful for
understanding mechanism of the heavy ion scattering in the field of the
deformed potential.

\subsection {Study of inelastic scattering}

Substituting the interaction phase (\ref{d13}) in (\ref{c5}) and
in the amplitude of scattering (\ref{c3}), we expand
$\exp(i\Phi_{int})$ leaving only the term of the first order in
$\beta_2$. Then, performing integration in (\ref{c1}) over
rotational angles $\{\Theta_i\}$ we obtain
%
\begin{equation}\label{e21}
f_{2\,0}(q)\,= \,-\,{k\over\sqrt{5}}\,\beta_2 \int_0^\infty \, bdb\, J_0(qb)\,
G_0(b)\,{\dis e}^{\dis i\Phi_0},
\end{equation}
%
\begin{equation}\label{e22}
f_{2\,2}(q)\,= \,{k\over\sqrt{5}}\,\beta_2 \int_0^\infty \, bdb\, J_2(qb)\,
G_2(b)\,{\dis e}^{\dis i\Phi_0},
\end{equation}

\begin{figure}[t]
\begin{center}
\includegraphics[ height = 3.in, width =
.95\linewidth]{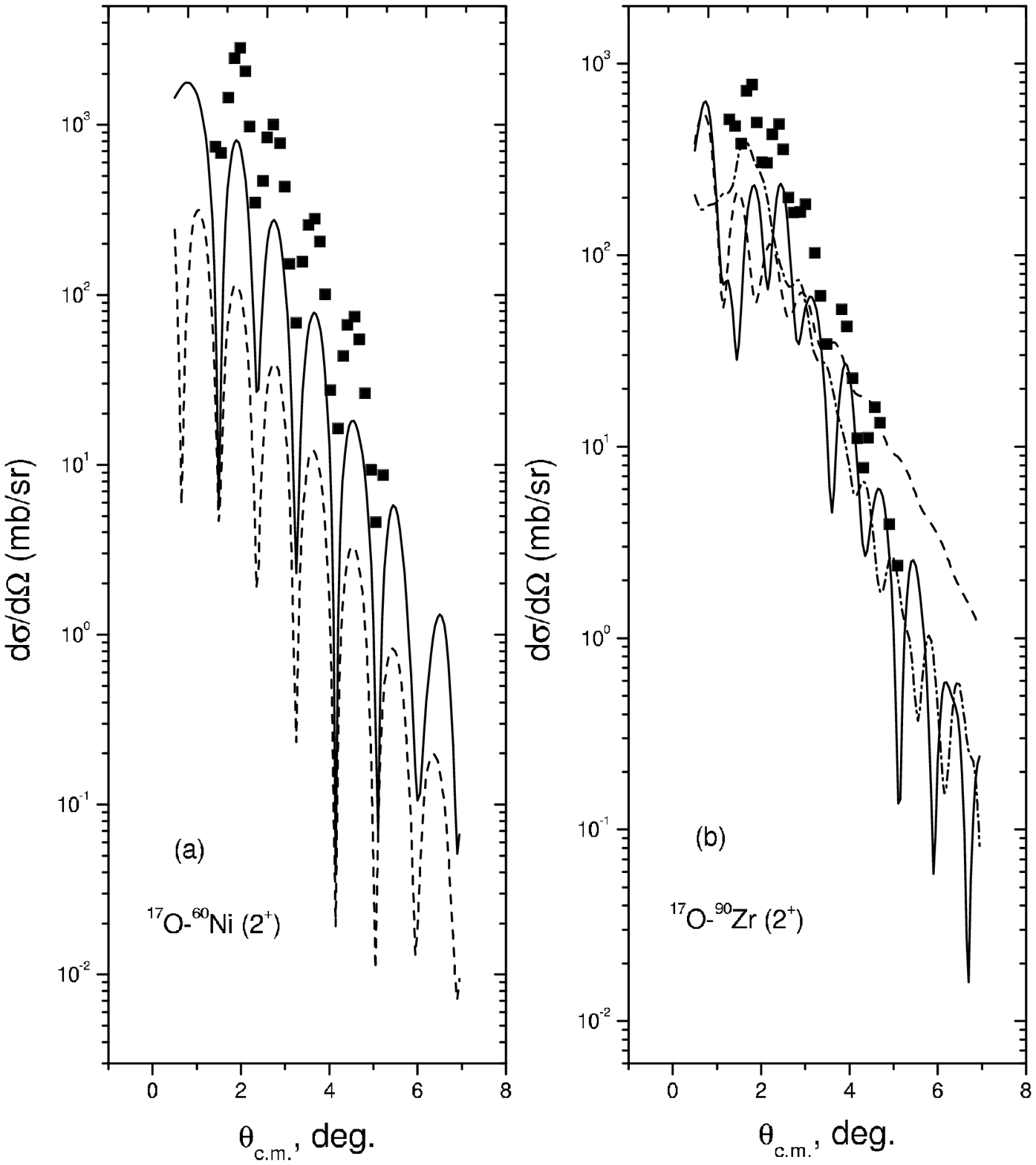}
\end{center}
{Fig.2. Inelastic scattering differential cross sections.
Experimental data are from \cite{Neto}. (a), case
$^{17}$O+${^{60}}$Ni: dashed curve - only M=0 amplitude is
included, solid curve shows - all terms of M=0 and M=$\pm 2$ are
included in the cross section (\ref{c7}). (b), case
$^{17}$O+${^{60}}$Zr: dashed and dot-dashed curves correspond to
the nuclear and Coulomb contributions in cross sections, solid
curve - the total contribution. }
\end{figure}

Figure 2(a) exhibits  differential cross sections (\ref{c7}) of inelastic
scattering of $^{17}$O+${^{60}}$Ni with excitation of the 2$^+$ rotational
state. The dashed curve corresponds to the case when only M=0 amplitude
$f_{2\,0}$ takes place, and the solid curve shows calculations with accounting
for all terms of M=0 and M=$\pm 2$ in the cross section (\ref{c7}). It is seen
that in inelastic scattering, mainly the M$\neq$0 amplitudes contribute to
cross sections. Then, in Fig.2(b), we exhibit contributions of the nuclear and
Coulomb interactions in inelastic scattering of $^{17}$O+${^{60}}$Zr (the
dashed and dot-dashed curves, respectively). It is seen that the both values
are of the same magnitudes and should be commonly included in calculations.
The further study shows that, for the lighter nucleus ${^{60}}$Ni,
nuclear interaction plays the more important role in comparison with the
Coulomb one, whereas for the $^{208}$Pb target-nucleus the Coulomb potential
influences greater on inelastic scattering than the nuclear one.

\begin{figure}[t]
\begin{center}
\includegraphics[ height = 5.in, width =
.95\linewidth]{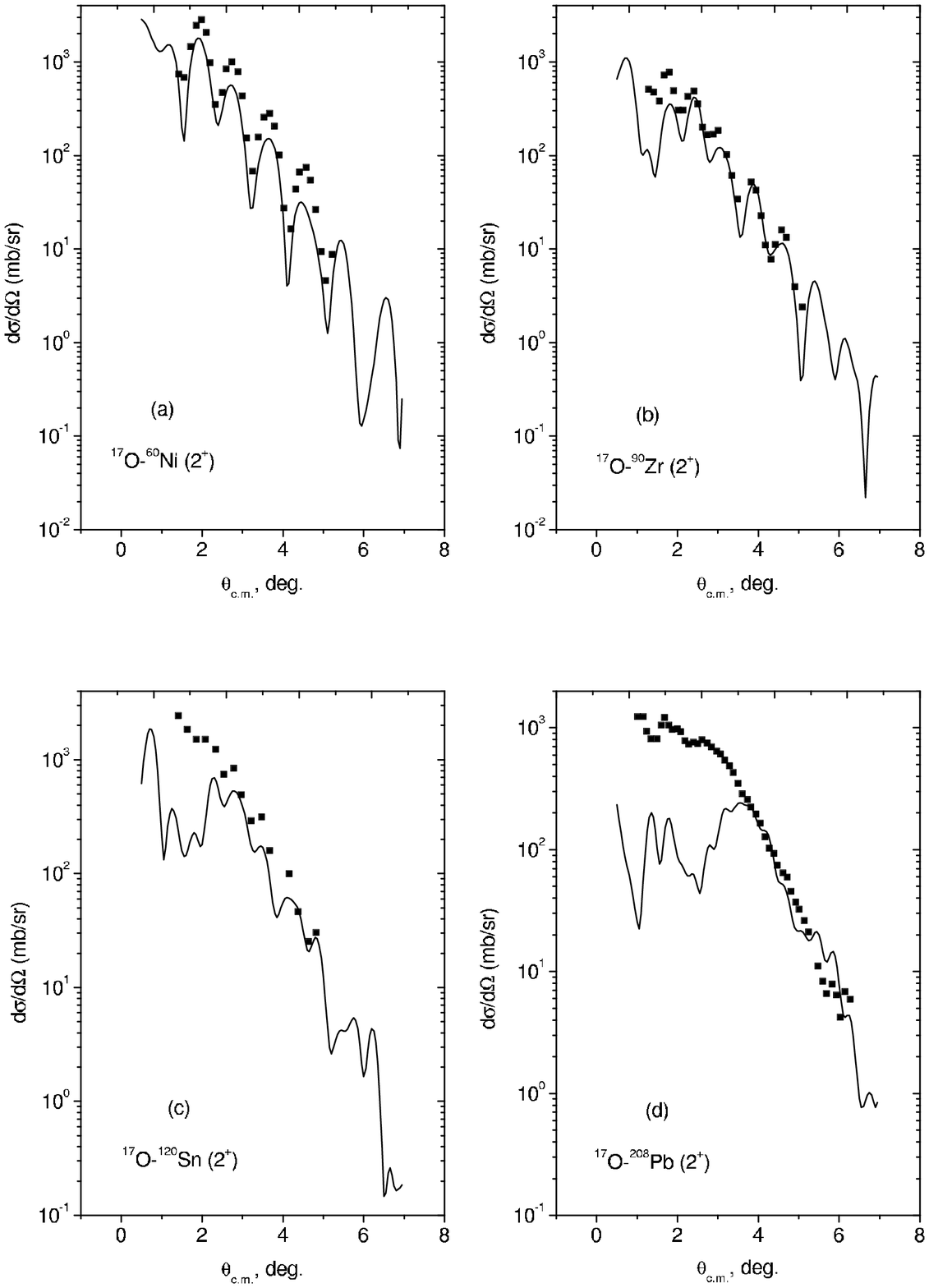}
\end{center}
{ Fig.3.  Comparison of the calculated inelastic scattering
differential cross sections with the experimental data from
\cite{Neto}. Deformation parameters see in the text.
 }
\end{figure}

In Fig.3, our calculations and the experimental data from \cite{Neto} are
shown of inelastic cross sections of the ${^{17}}$O heavy ions scattered on
the target-nuclei  $^{60}$Ni, $^{90}$Zr, $^{120}$Sn and $^{208}$Pb at
E$_{lab}$=1435 MeV with excitations of 2$^+$ states. These data were
also analyzed in \cite{Neto} utilizing the code ECIS \cite{Reinal} that
numerically solves the set of differential equations for coupled
elastic and inelastic channels. This latter theory is rather complicated to
study details of mechanism of a process, and when developing the obvious
adiabatic method and making comparisons with experimental data, we decided
to take the same parameters of potentials as they were obtained in \cite{Neto}.
Doing so, we wanted not only search details of the mechanism of scattering
but also to test an applicability of the suggested approach for high-energy
nucleus-nucleus scattering. To this aim, when comparing with experimental
data, we took the nuclear interaction potential $U^{(N)}_{int}$ in the same
form as in \cite{Neto}, namely, we exchange the derivative $du_{SF}/dR$ in
(\ref{d10}) by $-(r/R)du_{SF}/dr$. Also, deformation parameters were taken
separately for the nuclear and Coulomb potentials as $\beta_2^{(n)}$ and
$\beta_2^{(c)}$. They obey the equality $R\beta_2^{(n)}=R_C\beta_2^{(c)}$
and are distributed between the corresponding parts of the phase $\Phi_{int}$
(\ref{d13}), and are done as follows
$$
{^{60}}Ni:\qquad \beta_2^{(c)}\,=\,0.2067, \quad \beta_2^{(n)}\,=\,0.2356
$$
$$
{^{90}}Zr:\qquad \beta_2^{(c)}\,=\,0.0910, \quad
\beta_2^{(n)}\,=\,0.1000
$$
$$
{^{120}}Sn:\qquad \beta_2^{(c)}\,=\,0.1075, \quad \beta_2^{(n)}\,=\,0.1184
$$
$$
{^{208}}Pb:\qquad \beta_2^{(c)}\,=\,0.0544, \quad \beta_2^{(n)}\,=\,0.0597
$$
These deformation parameters correspond to nucleus-nucleus potentials
and, in general, they not to be just the same as for the target-nuclei.
However, in fact, for the three considered nuclei (an exception is for
${^{90}}$Zr) they give electric transition probabilities B(E2$\uparrow$)
in coincidence with the spectroscopic data.

One can see in Fig.3 that our calculations follow the experimental data.
Some discrepancies are seen at small angles, beginning from a slight
deviation for ${^{90}}$Zr to the larger one for ${^{208}}$Pb. Analisis
of these data in \cite{Neto} also show a little disagreement at small
angles for heavier nuclei. Small shortage of the value of our
calculations for ${^{60}}$Ni can be alleviated by a little increase of
a deformation parameter. In this connection we remind that in the coupled
channel method the potential and deformation parameters are adjusted to
experimental data self-consistently, and therefore the obtained parameters
of spherically symmetrical potentials do depend on deformation parameters
implicitly. So, in our adiabatic method, when we took these potentials
to construct their deformation admixtures, the problem of the double
accounting for deformation is really arise. In fact, it is obvious one
that in the framework of our method one can adjust parameters so that to
explain experimental data significantly better.

\section { Conclusions}
\begin{enumerate}
\itemsep -1mm
\item
Firsty, we conclude that the high-energy approach together with the
adiabatic method, when one takes into account intrinsic collective
excitations, occurs proved to be a suitable one to study mechanism of
scattering at energies in tens and higher Mev/nucleon and to analyze
corresponding experimental data on elastic and inelastic differential
cross sections.
\item
In elastic scattering, virtual excitations of collective states occur
weak and reveal themselves only at comparably large angles of scattering.
This effect can be simply accounted for if one adds to the ordinary
amplitute with the spherically symmetrical optical potential, the part
proportional to the second power of a deformation parameter.
\item
In the case of quadrupole excitations, contributions of amplitudes with
non-zero magnetic quantum numbers M=$\pm 2$ play the desicive role as
compared to the M=0 component. This is understandable if one reminds that
at high energies, the mechanism of scattering is mainly realized on the
plane perpendicular to the stright ahead trajectory of motion, where the
$Y_{2\,2}$ spherical harmonics are revealed prefarably.
\item
For heavy ions the Coulomb forces become comparable with the nuclear one
in the peripheral region of interactions. This is a reason why they
contribute to excitations of collective states fairly strongly and
give important effect on inelastic scattering process.
\item In conclusion we note, that the employed theory of inelastic scattering
where admixtures to interactions due to the deformation is accounted for by
inclusions of terms of the first order of $\beta_2$, does not distinguish the
rotational or vibrational nature of the 2$^+$ excited states. In the case of
vibrational states, $(\beta_2)^2$ has a meaning the middle squared value of dynamical
deformations of a surface.
\end{enumerate}

}

\begin{thebibliography}{99}
\itemsep -1mm
\bibitem{D1}
~S.I.Drozdov, JETPh {\bf 28}, 736 (1955).
\bibitem{I1}
~E.V.Inopin, JETPh {\bf 31}, 901 (1956).
\bibitem{B1}
~J.S.Blair, Phys. Rev. {\bf 115}, 928 (1959).
\bibitem{Ber1}
~Yu.A.Berezhnoy, A.P.Soznik, Yad. Fiz. {\bf 19}, 813 (1974).
\bibitem{Star1}
~V.E.Starodubsky, Nucl. Phys. {\bf A 219}, 525 (1974).
\bibitem{Gla}
~R.J.Glauber,~{\it Lectures in Theoretical Physics} (N.Y.: Interscience,
1959. P.315).
\bibitem{Sit}
~A.G.Sitenko, Ukr. Fiz. Zhurn. {\bf 4}, 152 (1959) ({\it in
Russian}).
\bibitem{LP}
~V.K.Lukyanov, I.Zh.Petkov, Izvest. Akad. Nauk SSSR Ser. Fiz. {\bf
29}, 823 (1965).
\bibitem{ISh}
~E.V.Inopin, A.V.Shebeko, JETPh. {\bf 51}, 1761 (1966).
\bibitem{FG}
~G.F{\"a}ldt, R.Glauber, Phys. Rev. {\bf C 42}, 395 (1990).
\bibitem{LZ2}
~V.K.Lukyanov, E.V.Zemlyanaya, Int. J. Mod. Phys. {\bf E 10}, 169
(2001).
\bibitem{LSZ}
~V.K.Lukyanov, B.S{\l}owi{\'n}ski, E.V.Zemlyanaya, Yad. Fiz. {\bf
64}, 1349 (2001); Phys. At. Nucl., {\bf 64}, 1273 (2001).
\bibitem{LZ}
~V.K.Lukyanov, E.V.Zemlyanaya, J. Phys. {\bf G 26}, 357 (2000).
\bibitem{Neto}
~R.L.Neto e.a., Nucl. Phys. {\bf A 560}, 357 (1993).
\bibitem{Reinal}
~J.Reinal, Phys. Rev. {\bf C 23}, 2571 (1981).
\end{thebibliography}
\end{document}